\documentclass[aps,twocolumn,showpacs]{revtex4}
\usepackage{graphicx,hyperref,subfigure,color}
\usepackage{epstopdf}
\begin{document}
\title{Microrheology in tumbling nematics of 8CB liquid crystals}
\author{Balaji Yendeti,  Ashok Vudaygiri\footnote{corresponding author: email avsp@uohyd.ernet.in}, }
\affiliation{School of Physics, University of Hyderabad, Hyderabad, 500046 {\bf India}}
\begin{abstract}
Particle tracking passive microrheology in 8CB liquid crystals is used to redefine the precessional motion of the orientation of nematic director  in liquid crystals. Physical origin of tumbling director in presence of presmectic clusters under zero shear conditions is discussed. Different structural properties (pure nematic phase, presmectic(smectic C and smectic A clusters)) were differentiated with characteristic dependence of $G'$ on $\omega$ in the nematic phase of 8CB liquid crystals. Also, dynamic viscosity is observed  with a cross over between parallel and perpendicular components as the smectic A phase is approached. 
\end{abstract}  
\maketitle
\section{introduction}

Complex fluids driven to induce flow shows the characteristics of irregular intervals of solid like behavior followed by stress relaxations due to rearrangements of its constituents. Dynamics of complex fluids are non-linear in nature.  Many of these condensed matter systems are either time dependent or frequency dependent. Dynamic properties of the condensed matter systems are mostly harmonic oscillator like modes. Information about the dissipation of harmonic oscillators  energy  into incoherent degrees of freedom is reflected in the sign of the dissipative forces that leads to decay of amplitude of motion in time\cite{P.M.Chaikin}. On this note, measuring passive microrheological parameters using single particle tracking method will lead to understanding the dynamics of flow behavior in complexfluids.
   
In shear rheology experiments of liquid crystals, Ericksen-Leslie-Pondou(ELP) theory was used in understanding various regimes in the nematic phase of liquid crystals.  In nematic phase of liquid crystals, the average orientation of molecules is determined by the nematic director $(\hat{\textbf{n}})$\cite{P.G.de Gennes}. This director field is coupled to the  velocity field or flow field $\textbf{V(r,t)} $. The dissipation of energy  in liquid crystals strongly depends on the microscopic constituent molecules alignment and orientation. According to ELP theory, the direction of orientation of $(\hat{\textbf{n}})$ characterizes the nematic stress tensor $(\alpha_{i})$ Where $i=1$ to $6$. Miesowicz viscosities $\eta_{1},\eta_{2},\eta_{3}$ depends on $(\alpha_{i})$. The sign of the viscosity coefficient $\alpha_{3}$ determines the flow aligning behavior  of the NLC. The sign of $\alpha_{2}$ is always negative in case of nematic liquid crystals, so, if $\alpha_{2}$,$\alpha_{3}$ satisfy a condition, such that $\frac{\alpha_{2}}{\alpha_{3}}>0$, a flow alignment takes place, with director orientation  at an angle $\tan \theta =\sqrt{ \frac{\alpha_{2}}{\alpha_{3}}}$. In standard liquid crystals like 5CB, with sequence of phases as crystal(cr), nematic(N) and Isotropic(I) fluid, then flow alignment occurs with the director at a characteristic angle with respect to the flow direction even at low shear fields, in the whole nematic phase\cite{K. Negita}. For $\alpha_{3}>0$, the director is forced to rotate continuously by the hydrodynamic torques even at low shear field.  The director rotation angle $\theta$ will always increase with strain  and an indefinite oscillatory response is predicted\cite{D.F.Gu}. Jahnig et.al.\cite{Jahnig} observed that, if a liquid crystal possess a phase sequence as crystal(cr), smectic A(Sm-A), nematic(N), and Isotropic phases, then it will have positive $\alpha_{3}$, in its nematic phase near the Sm-A to N transition temperature, since $\alpha_{3}$, has a positive divergence at $T_{SmA-N}$. In 8CB liquid crystals, phase sequence occurs as cry-Sm-A-N-I, so for $\alpha_{3}>0$, nematic phase  is of the tumbling type in these liquid crystlas. It is well known that there exists several structures in its nematic regime of 8CB liquid crystals 
in addition to phase transitions \cite{K. Negita, K. Negita-1, C. R. Safinya} In particular, the orientation or 
the isotropic and anisotropic precessional motions of the director around the velocity or the flow gradient directions determines the phases in the nematic regime.    From theoretical\cite{G. Marrucci},\cite{W. H. Han}, and experimental  study \cite{D.F.Gu}  of variation of apparent viscosity with shear rate and temperature in 8CB liquid crystals, it was observed that there exists tumbling motions of the director that dominates the whole nematic regime. These results  also showed the periodicity of the tumbling oscillations to be a strong function of temperature.  This  tumbling motion in the nematic phase of 8CB liquid crystals was explained more precisely by K.Negita et.al.,\cite{K. Negita-1}.  Various structures are mentioned in their paper with different notations. Here we follow the same notations in representing these structures. It can be  briefly recollected  as: (i) In the temperature region just below the N-I transition point,
   a flow alignment of the director occurs with its direction near the flow direction(y-axis). In other words, the liquid crystals in this regime behave as  flow aligning nematogens. (ii) Below this  temperature, 
  $ \frac{\alpha_{2}}{\alpha_{3}} $ becomes negative, making the flow alignment of the director impossible. The structures formed in this regime are denoted $a-b, a_{m},a_{s},a(b)$ and $a_{c}$ with their stable regions depending on the temperature and shear rate.  The $a-b$ structure is composed of coexistent $a$ and $b$ structures. Further below this  temperature, precessional motions around the x axis are induced, leading to $a_{m}, a_{s}, a(b)$ and $ a_{c}$ structures. These  precessional motions are characterized by
   $ \frac{n_{y}(t)^{2}}{n^{2}_{y0}}
    + \frac{n_{z}(t)^{2}}{n^{2}_{z0}} = 1 $.  Here $n_{y}(t) = n_{y0} cos(\omega_{0}t)$ and $n_{z}(t)= n_{z0} sin(\omega_{0}t) $ are components of the director $ n(t) = (n_{x}(t),n_{y}(t),n_{z}(t))$ precessing around the x-axis with an angular frequency of $\omega_{0} = [\frac{\dot {\gamma}^{2}(-\alpha_{2}\alpha_{3}^{R})}{\gamma^{2}_{1}}]^{\frac{1}{2}} $. $\dot{\gamma}$ is the shear rate, $ \alpha_{3}^{R} $ is the renormalized viscosity coefficient of $\alpha_{3} $ including the effect of the $S_{A}$ fluctuations, and $\gamma_{1} = \alpha_{3}-\alpha_{2}$. Further, the precessional motions in each structure are explained as: 
			\\1)$a_{m}$ strucure: anisotropic precession with larger amplitude in the y-direction with $(n_{y0}>n_{z0})$.
			  2)$a_{s}$ structure: isotropic precession with equal amplitude in the y and z direction $(n_{y0}= n_{z0} $.
			  3)$a(b)$ structure: anisotropic precession with larger amplitude in the z direction $(n_{y0} < n_{z0} $.
			  4)$a_{c}$ structure: anisotropic precession with its  larger amplitude of motion deflected along the z-axis $(n_{y0}<< n_{z0}) $
			  and in the $S_{A}$ phase, the formation of the layer structure makes the precessional motion impossible.
			  
			  All the above description of structural regimes involve the  ELP theory and critical nematodynamics in bulk rheology and X-ray diffraction measurements. In this paper we present instead a real space in-situ measurement using Brownian motion of microsphere. The diameter of the microsphere is about the same size as that of the structures formed by the orientations of director $(\hat{\textbf{n}})$. This measurement of Brownian motion in tumbling nematic phase is first of its kind. The microsphere motion among these structures within liquid crystals is in analogy to motion in a harmonic trap, because of the defect formed around the microsphere. In such a case, the dissipative force describes the average effect on the Brownian motion of bead in liquid crystals.  Hence it is important to measure $G^{'}$ and $G^{''}$ with out using ELP theory of shear rheology in 8CB liquid crystals. 
			  
\section{The Experiment}
 Silica microshperes of diameter 0.98 $ \mu m $  are coated with Octadecyldimethy(3-trimethoxysilylpropyl) ammonium chloride(DMOAP, Sigma Aldrich) for  a  specific molecular alignment \cite{M.Skarabot} and dispersed in the 8CB (4- Cyano  $4^{'}$ - Octyl biphenyl) liquid crystals(Sigma Aldirich).  Liquid crystal cell is made of two coverslips coated with AL-1254 and rubbed in opposite directions to achieve planar alignment. The measured thickness of the sample cell is 23$ \mu m $. Observed 8CB liquid crystals phase transitions(cooling) are Isotropic at $41.5^{\circ} C$ to Nematic at $ 34.8^{\circ} C$ to Sm-A. The resolution of the temperature controller (Instek, mk 1000) is $\pm 0.01^{\circ} C$.    Brownian motion of an isolated particle is tracked with an appropriate computer program with in an inverted polarizing microscope (Nikon Eclipse, Ti-U) using a 100X objective. The experimental measurements of Brownian motion is very similar to our previous paper \cite{dhara}. The background  for calculating microrheological parameters like G$'$ and G$''$ from the analysis of Brownian motion  is from \cite{Balaji},\cite{T.G.Mason-2}, \cite{T.G.Mason}.  The Brownian motion of an isolated micro sphere is video recorded for about 5 minutes and analyzed. Measurements are repeated with in the same and different cells.
 
\section{Results and discussion}
When a microsphere of  0.98$\mu m$ diameter moves in the nematic phase of liquid crystal, the fluctuating director $(\hat{\textbf{n}})$  displaces or biases the thermal motion of the microsphere. Suppose, a fluctuative director$(\hat{\textbf{n}})$ displaces the sphere to its left by a distance $\Delta$\textbf{x}, this displacement temporarily increases the elastic energy density on the left side. So, there occurs a difference in the elastic energy densities between left and right side. This leads to a restoring elastic force F = -K $\Delta$\textbf{x}/d. Here, $d$ is the diameter of the sphere. This force slows down the diffusion of the microsphere \cite{T. Turiv}.
   If the microsphere moves through structures  of the order of diameter of sphere($\sim 1\mu$ m), then the thermal motion of the sphere is constrained due to these structures. This leads to an anomalous Brownian motion of the microsphere.
\begin{figure}
\subfigure{\includegraphics[scale=0.3]{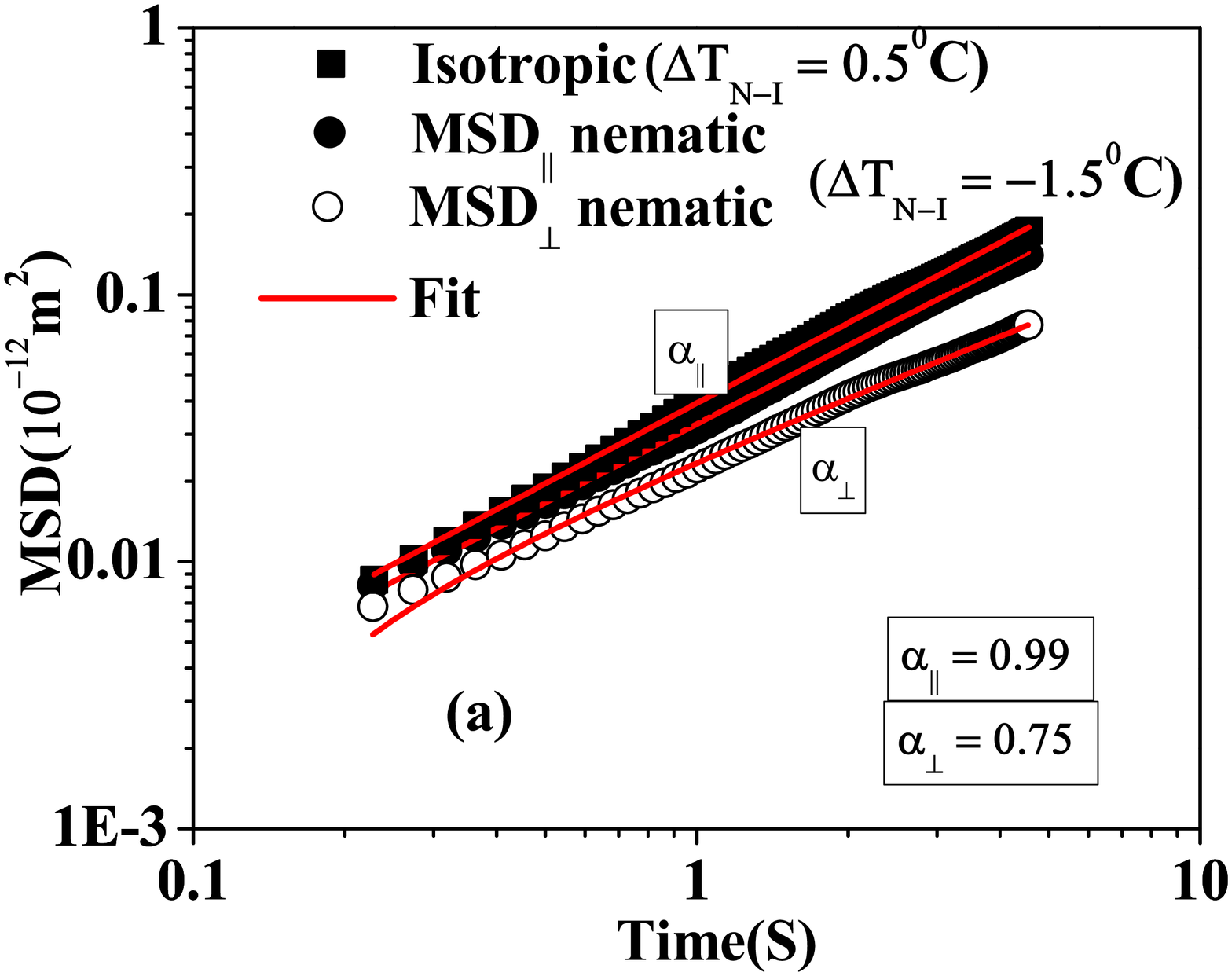}\label{8CBMSD-1}}
\subfigure{\includegraphics[scale=0.3]{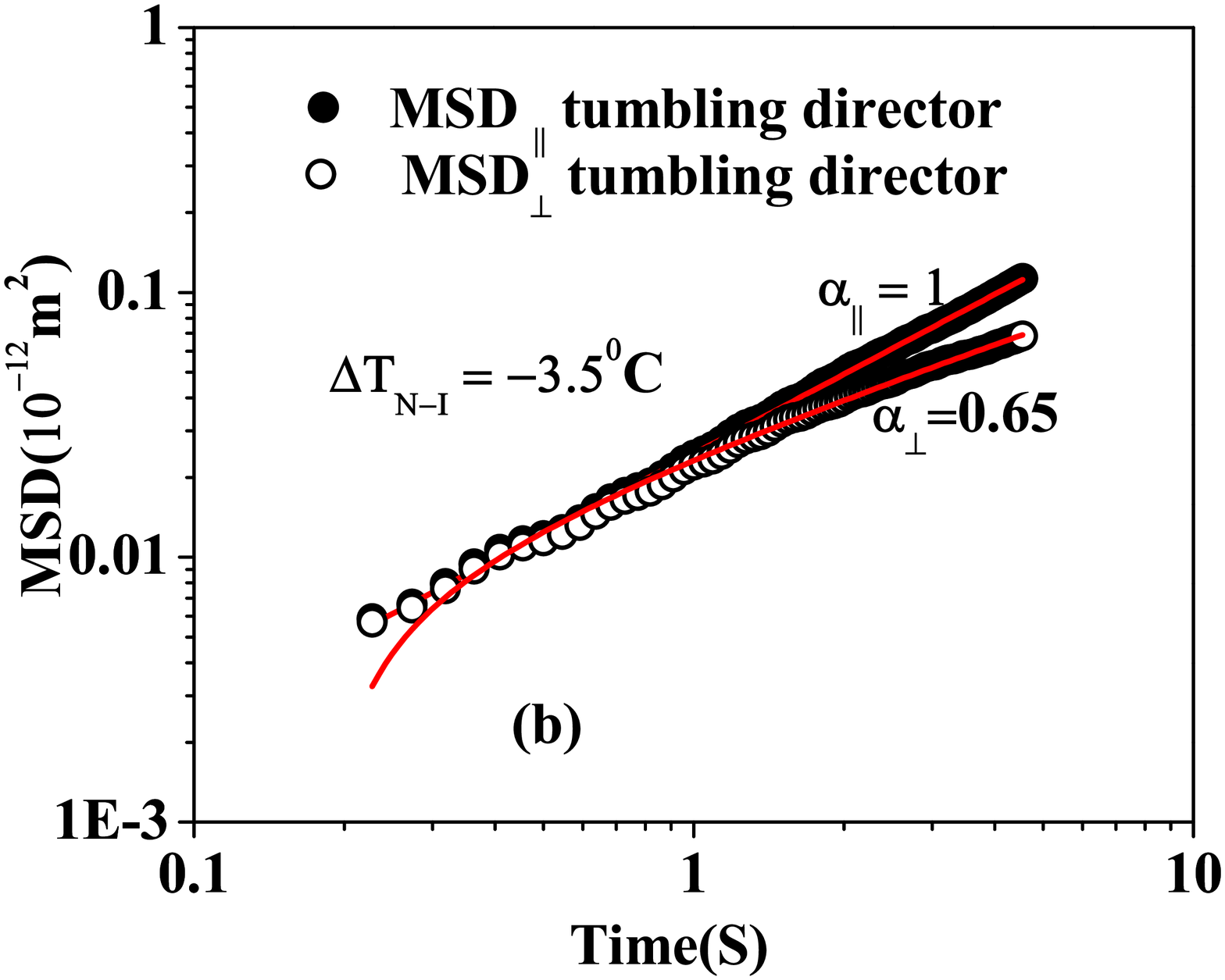}\label{8CBMSD-2}}
\caption{MSD vs time for (a) Isotropic(solid square) $(\Delta T_{N-I} = + 0.5^{\circ}C)$ and nematic phase(closed and open circles for parallel and perpendicular MSD)  $(\Delta T_{N-I} = - 1.5^{\circ}C)$  and (b) In case of tumbling nematic director(closed and open circles for parallel and perpendicular MSD) at $(\Delta T_{N-I} = - 3.5^{\circ}C)$  of 8CB liquid crystals. solid line(red colour online) represents the power law fit.}

\end{figure}

 In isotropic fluids, the mean square displacement (MSD) of the Brownian motion of a spherical particle is linear in time $\tau$. But in case of complex fluids such as entagled actin filament networks etc., the Brownian motion is anomalous, mostly due to the infrequent large jumps that the particle makes between distinct pores in the network\cite{I.Y.Wong}.  Hence the MSD($\langle\Delta{r^{2}(t)}\rangle$) follows a power law behaviour with $\langle\Delta{r^{2}(t)}\rangle\propto\tau^{\alpha}$. When  $0<\alpha<1$ the nature is sub-diffusive  and when  $\alpha>1$ it is  super-diffusive, over a lag time $\tau$. Recently, sub and super-diffusive motion in 5CB liquid crystals have been reported by measuring time variation of $\left( \langle \Delta r^{2}(t)\rangle \right)$ \cite{T. Turiv}.  Here, we present the MSD results from the thermal motion of the colloidal particle in 8CB liquid crystals, and show how precessional motions of the director and the collective molecular reorientations affect the Brownian motion.
  In figure \ref{8CBMSD-1}, variation of MSD with time lag  at $\Delta T = 0.5^{0}C$, $\Delta T = -1.5^{\circ}C $ are presented.    In isotropic regime, at $\Delta T = 0.5^{0}C$, we observed linear increase in the value of $\alpha$ with variation in time lag $\tau$.  
  When temperature is further decreased  to $\Delta T = -1.5^{\circ}C $, an anisotropy in the MSD Vs time can be seen as shown in fig.\ref{8CBMSD-1}. In the log-log plot, the parallel component possesses linear behavior with MSD vs time and fits well with linearly increasing curve whereas the  perpendicular component shows sub diffusive behavior  and fits well with power law rise in time and the value of $\alpha = 0.77$. Further, at $\Delta T = -3.5^{0}C$, the parallel component of $\left( \langle\Delta{r^{2}(t)}\rangle\right)$ show linear increase with increase in time and perpendicular component exhibits non-linear increase in $\alpha$ value as shown in figure \ref{8CBMSD-2}, and shows sub diffusive behavior with $\alpha$ value being equal to 0.65. In order to further investigate sub diffusive properties of Brownian motion of colloidal particle  in 8CB liquid crystals, we computed microrheological parameters $G'$ and  $G''$.
  
  
  	Many of the condensed matter systems are either time dependent or frequency dependent. Dynamic properties of the condensed matter systems are mostly harmonic oscillator like modes. Information about the dissipation of harmonic oscillators  energy  into incoherent degrees of freedom is reflected in the sign of the dissipative forces that leads to decay of amplitude of motion in time.

  From the measured thermal motion of the colloidal particle,    viscoelastic moduli (storage modulus$(G')$ and  loss modulus $(G'')$),  can be computed. G$'$ and G$''$ represent the fraction of energy induced by the  deformation imposed on the material. A fraction of energy is lost by the viscous dissipation in case of loss modulus(G$''$), and, is stored elastically in case of storage modulus(G$'$). G$'$ and G$''$ were computed using the  equations given in Ref.\cite{T.G.Mason-2}, given below,  as a function of rate of deformation $\omega$. 
By expanding the equation of motion of Mean square displacement around $t=1/\omega$, the translational parameter  $\beta$ is defined as \cite{T.G.Mason-2}
									
\begin{equation}
\beta(\omega)=\left. \frac{\langle d \ln \Delta r^2(t)\rangle}{d \ln t} \right|_{t=\frac{1}{\omega}}
\end{equation}

and the Fourier transform of this MSD equation would lead to give G$'$($\omega$) and G$''$($\omega$) as

\begin{eqnarray}
G'(\omega)&=&\vert G^{*}(\omega)\vert \cos\left(\frac{\pi\beta(\omega)}{2}\right) \cr
\cr
G''(\omega)& =& \vert G^{*}(\omega)\vert \sin\left(\frac{\pi\beta(\omega)}{2}\right)
\end{eqnarray}
Where $\vert G^{*}(\omega)\vert=K_{B}T/\pi a \langle \Delta r^2(\frac{1}{\omega})\rangle \Gamma[1+\beta(\omega)]$, and `$a$' the radius of the probing bead.

\begin{figure}
\includegraphics[scale=0.33]{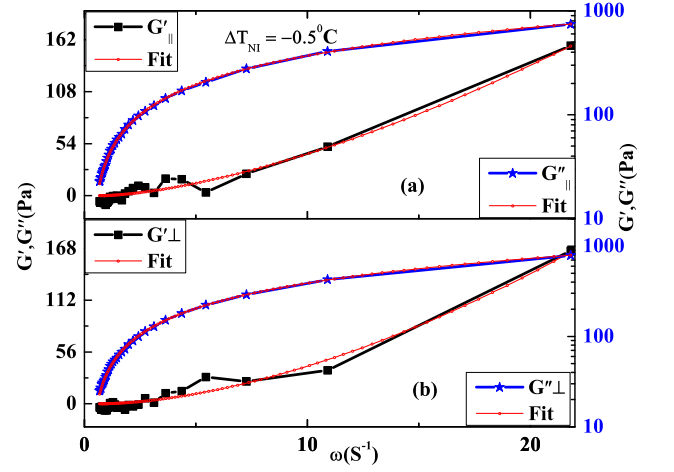}
\caption{In nematic phase of 8CB liquid crystals, at $(\Delta T_{N-I} = - 0.5^{\circ}C)$, $G^{'}_{\|}$ and $G^{''}_{\|}$  represent the parallel components of $G^{'}$ and $G^{''}$. Here, $G^{'}$ is represented by  line with solid square(black colour online) and $G^{''}$ is represented by line with star(blue colour online).(b) $G^{'}_{\bot}$ and $G^{''}_{\bot}$ represents perpendicular components of $G^{'}$ and $G^{''}$. Thin line(red colour online) with a small open circle shows fitting.}
\label{8CB G para tumb}
\end{figure}

\begin{figure}
\includegraphics[scale=0.33]{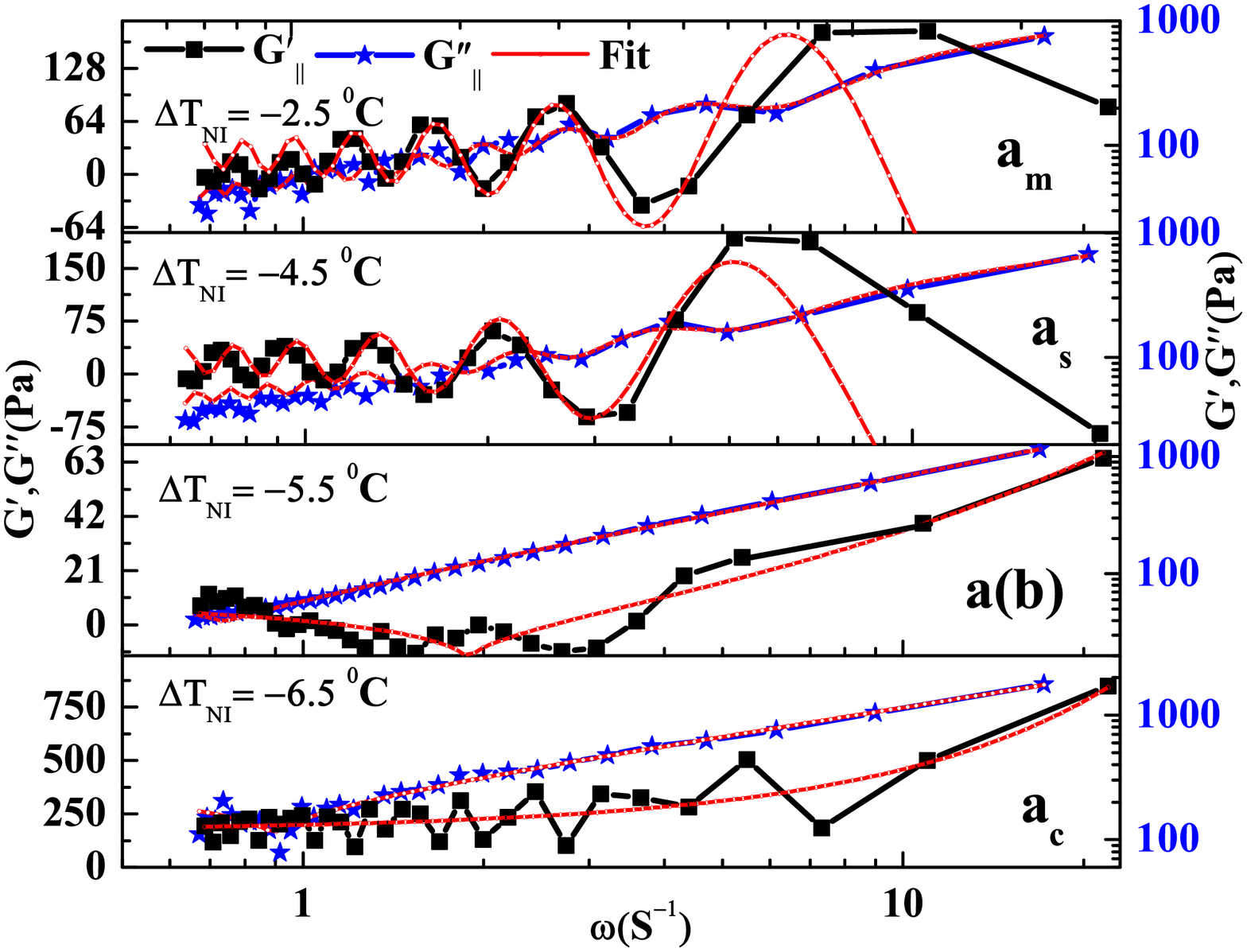}
\caption{In nematic phase of 8CB liquid crystals, parallel components of elastic ($G^{'}_{\|}$) and viscous modulus ($G^{''}_{\|}$) vs frequency ($\omega$) of different regimes($a_{m},a_{s},a(b),a_{c}$). Here, $G^{'}$ is represented by  line with solid square(black colour online) and $G^{''}$ is represented by line with star(blue colour online). Thin line(red colour online) with a small open circle shows fitting.}. 
\label{8CB G para tumb}
\end{figure}
\begin{figure}
\includegraphics[scale=0.33]{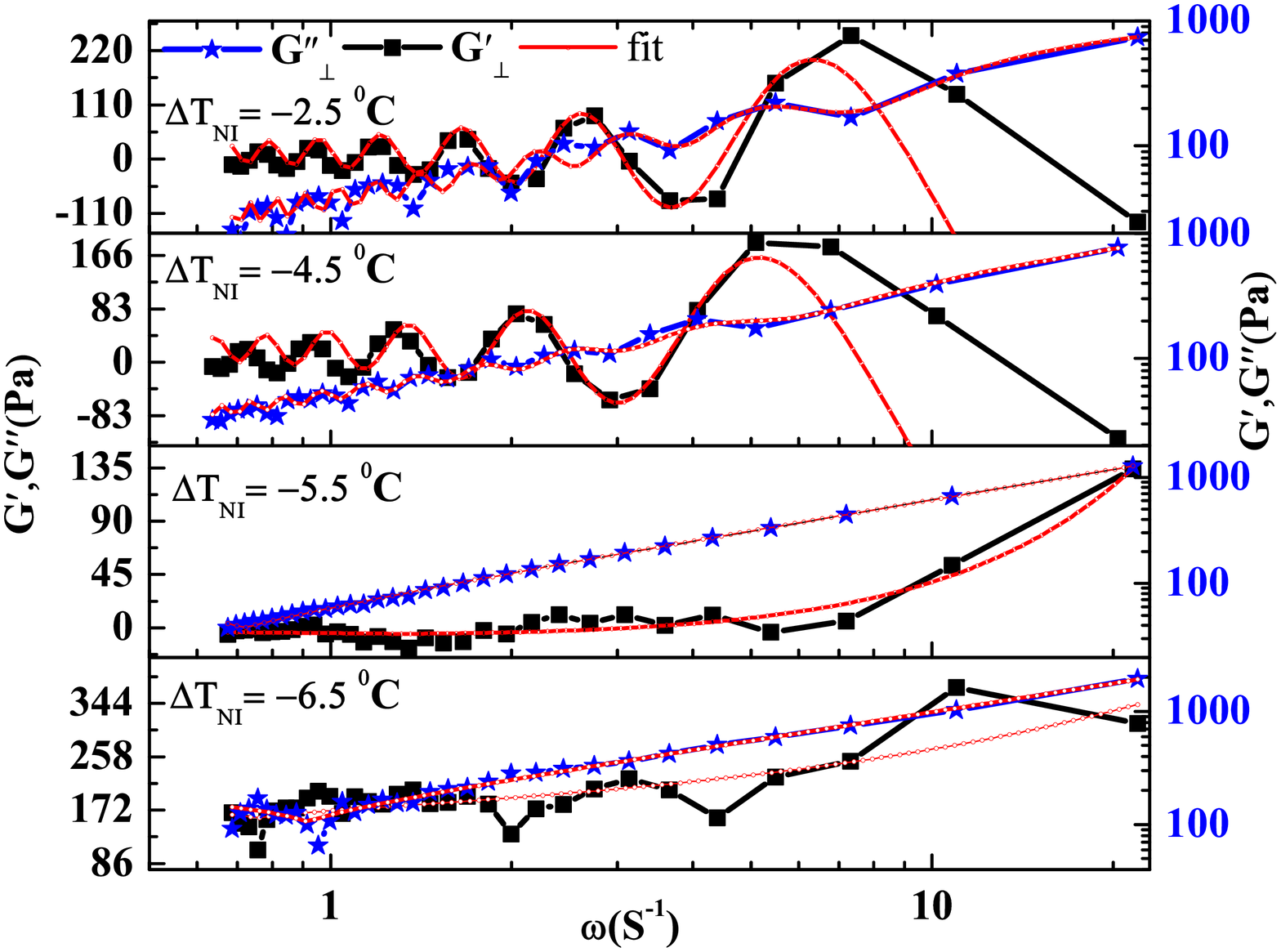}
\caption{In nematic phase of 8CB liquid crystals, perpendicular components of elastic ($G^{'}_{\bot}$) and viscous modulus ($G^{''}_{\bot}$) vs frequency ($\omega$) of different regimes($a_{m},a_{s},a(b),a_{c}$). Here, $G^{'}$ is represented by  line with solid square(black colour online) and $G^{''}$ is represented by line with star(blue colour online). Thin line(red colour online) with a small open circle shows fitting.}
\label{8CB G perp tumb}
\end{figure}

Here, we present the microrheological properties elastic modulus $G'$, viscous modulus $G''$ in the nematic phase of 8CB liquid crystals. In liquid crystals, the average orientation of liquid crystal molecules is determined by the director $(\hat{\textbf{n}})$. Hence, the effect of collective molecular reorientations on Brownian motion of colloids in nematic liquid crystals is determined by the direction of orientation of $(\hat{\textbf{n}})$.  Just below nematic to isotropic transition temperature, in $b^{'}$ regime, at $(\Delta T_{N-I} = - 0.2^{\circ}C)$, a flow alignment of the nematic director $(\hat{\textbf{n}})$ occurs with its direction with a small angle near the flow direction(Y-axis). Such flow alignment also occurs in 5CB liquid crystals. In figure \ref{G in b'}, 
the power law dependence of frequency of $G'(\omega)$,    $G''(\omega)$ in both parallel and perpendicular directions to the orienatation of $(\hat{\textbf{n}})$ are  shown with their corresponding fits. From the curve fitting, it represents 
\begin{eqnarray}
 G'(\omega)_{\|}\sim \omega^{1.5} \cr
G'(\omega)_{\perp}\sim \omega^{1.6} \cr
 G''(\omega)_{\parallel,\perp} \sim \omega^{0.88}
 \label{G in b'}
\end{eqnarray}

According to the work of  C.R.Safinya et.al.\cite{C. R. Safinya}, the orientational  state of  $(\hat{\textbf{n}})$ is determined by the viscous, elastic and fluctuation torques due to fluctuation smectic clusters acting on it.  In shear rheology, the elastic torques are normally neglected. However, on lowering the temperature towards $T_{N\longrightarrow Sm-A}$,  the shear flow tends to tilt the smectic cluster layers formed.  This changes the layer spacing and gives rise to the restoring force. We have been able to observe the effect of the tumbling of the nematic director on the elastic and storage modulii, even in absence of an external shear. In other words, the tumbling need not be a  be a consequence of external shear on the smectic cluster layers and we propose an alternative explanation.


\subsection{Physical origin of tumbling nematic director :}
Below $a-b$ regime, tumbling of  $(\hat{\textbf{n}})$  is mainly because of presmectic translations. Here, the reason for the tumbling effects of director $(\hat{\textbf{n}})$  under zero shear conditions can be expected in two ways:  1) The molecular layers formed in these presmectic(Smectic A) translations slide freely one over the other resulting in shear effects. A small flow of molecules along the director orientation and the shear effects among presmectic layers act perpendicular to each other resulting in a  superposition of each other.  This superposition  leads to tilt the clusters of molecules in the presmectic (Smectic C) layers.  Which changes the layer spacing and gives rise to elastic torque \cite{T. Turiv}.   2) As the temperature decreases,   these presmectic layers gets rearranged and hence  exists compression on to the central portion of the liquid crystals.  Because of the quasistaic compression on to the liquid crystal layers in the central portion of the sample cell, this could necessarily affect the equilibrium structure in two different ways. $a)$ The gap between the layers inside the central portion of the sample cell would either be increased or decreased in order to adopt to the variations of the gap size. $ b)$ In order to relax from the elastic stress of the layers, the radius of the dislocation loops would be either be increased or decreased\cite{P.C.Martin}.  In simple words,  the layers that are close to the wall boundaries, gets rearranged with the changes in temperature. In process, the elastic energy density increases between layers. In single particle tracking experiments in liquid crystals, the change in elastic energy density between layers and central portion of the cell leads to restoring force on the microsphere. 

In liquid crystals, the average orientation of liquid crystal molecules is determined by the director $(\hat{\textbf{n}})$. Hence, the effect of collective molecular reorientations on Brownian motion of colloids in nematic liquid crystals is determined by the direction of orientation of $(\hat{\textbf{n}})$. In 8CB liquid crystals cell, the parallel and perpendicular orientations to $(\hat{\textbf{n}})$ represent Y(flow) and Z(flow gradient) directions respectively. As the nematic director does precessional motion to Y and Z directions, consequently, the thermal motion of the microsphere gets affected or biased by the precessional motion of the director. Hence, in the amplitude and periodicity of the microrheological properties, the effect of the precesional motion of the director $(\hat{\textbf{n}})$ is profound.  Amplitude and periodicity of $G'(\omega)_{\|}$  in figure \ref{8CB G para tumb} show strong temperature dependence. These characteristics of $G'(\omega)_{\|}$  are   analogus to  the apparent viscosity $\eta$ vs shear strain  as in a paper by D.F. Gu et.al.\cite{D.F.Gu}. Amplitude of harmonics increase first from $a_{m}$ regime to $a_{s}$ regime and then decreases rapidly in $a(b)$ regime and then again increases in $a_{c}$ regime more steeply. But, from figure \ref{8CB G perp tumb} amplitude of harmonics of $G'(\omega)_{\bot}$  decreases from $a_{m}$ regime to $a(b)$ regime and then increases in  $a_{c}$ regime. The periodicity of the harmonics are present in $a_{m}$  and $a_{s}$ regimes of both $G'(\omega)_{\|}$ and $G'(\omega)_{\bot}$.  In $a_{m}$ regime, as  the components of the director show $(n_{y0} > n_{z0})$. Here, $G'(\omega)_{\|}$ show anisotropy with $G'(\omega)_{\bot}$ and $G'(\omega)_{\|} > G'(\omega)_{\bot}$. This represents, anisotropy in $G'(\omega)_{\|}$ and  $G'(\omega)_{\bot}$ is clearly a consequence of amplitude of precessional motion of the director $(\hat{\textbf{n}})$ being larger in the Y-direction than in the Z-direction. In $a_{s}$ regime, $G'(\omega)_{\|} = G'(\omega)_{\bot}$. This is a consequence of isotropic precession of  $(\hat{\textbf{n}})$  with equal amplitude in both Y and Z-directions $(n_{y0} = n_{z0})$. In $a(b)$ regime, $G'(\omega)_{\|} < G'(\omega)_{\bot}$. This is a consequence of  anisotropic precession with lesser amplitude in the Y-direction than in the Z-direction $(n_{y0} < n_{z0})$.  In  $a_{c}$ regime, both  $G'(\omega)_{\|}$ and  $G'(\omega)_{\bot}$ does show larger values. Here,  $G'(\omega)_{\|} \gg G'(\omega)_{\bot}$. This is in contrast to the  anisotropic precession of  $(\hat{\textbf{n}})$,  with very large amplitude in the  Z-direction $(n_{y0} << n_{z0}) $.  This is actually a crossover between values of $G'(\omega)_{\|}$ and $G'(\omega)_{\bot}$.  Since $a_{c}$ regime is very close to nematic-smectic transition, smectic layers form temporary clusters.  Hence, the elastic energy density among these layers increases. Since, these temporary smectic layers are in the perpendicular direction(Z-axis), the difference in the elastic energy density between the central plane(Y-axis) and temporary smectic clusters in the Z-axis creates restoring elastic force on the microsphere. This slows down the diffusive motion of the microsphere along the parallel direction and it results in increasing the elastic or storage modulus $G'(\omega)_{\|}$.  

 Further, the dependence of elastic modulus $G'(\omega)$ on $\omega$ is measured directly by curve fitting. 
\begin{equation}
G'(\omega)_{\|,\bot} \propto \omega_{\|,\bot}*sin\left( \frac{\gamma_{p}}{\omega_{\|,\bot}}\right) 
\label{coupled}
\end{equation}
 The above equation \ref{coupled} represents sinusoidal distribution of $G'(\omega)$ with respect to deformation frequency $\omega$ in $a_{m}$ and $a_{s}$ regimes. The positive and negative slopes in the distribution of  $G'(\omega)$ indicates the storage and dissipation of elastic modulus.  These positive and negative slopes of  $G'(\omega)$ are consequence  of  tilt in the clusters of molecules of pre smectic layers.  This tilt in these clusters of molecules of Smectic layers leads to Smectic C like clusters locally. Whence with this micro rheology method, discrimination between clusters of presmectic A and clusters of Smectic C is identified. 
 
   In $a(b)$ regime, $G'(\omega)$ vs $\omega$ followed power law dependence. 
\begin{equation}
G'(\omega)_{\|} \propto \omega_{\|}^{0.5} 
\label{G' para in a(b)}
\end{equation}
  and
  \begin{equation}
  G'(\omega)_{\bot} \propto  \omega_{\bot}^{1.3} 
  \label{G'perp in a(b)}
  \end{equation}
Also, in $a_{c}$ regime,
\begin{equation}
G'(\omega)_{\|} \propto  \omega_{\|}^{1.7} 
\label{G'para in ac}
\end{equation} 
 \begin{equation}
  G'(\omega)_{\bot} \propto \omega_{\bot}^{0.6}
  \label{G' perp in ac} 
  \end{equation}
 These equations \ref{G' para in a(b)},\ref{G'perp in a(b)},\ref{G'para in ac},\ref{G' perp in ac} show  anisotropy in power law dependence of  $G'(\omega)$ vs $\omega$  in both parallel and perpendicular directions. This $a(b)$ regime is known to be a combination of $a$ and $b$ orientations of liquid crystal molecules.  In $a(b)$ regime, the power law dependence of $\omega$ in $G'(\omega)_{\|}$ vs $\omega$ is similar to the power law dependence of $ G'(\omega)_{\bot}$ in $a_{c}$ regime.  The power law dependence of $ G'(\omega)_{\bot}$ is similar to power law dependence of $G'(\omega)$ vs $\omega$ in $b'$ phase \ref{G in b'}. This shows how the orientation of director is tumbling between Y(a-orientation) and Z(b-orientation)directions.  In $a_{c}$ regime, from equation \ref{G'para in ac} represents the precessional motion of amplitude of the director being large to the Z direction. Hence  the parallel and perpendicular compoents of   $G'(\omega)$ flip between each other.  In both $a(b)$ and $a_{c}$ regimes, the reduction in the power of either of the parallel or perpendicular components represent the presence of pre-smectic A clusters of molecules.  since the cell is rubbed for planar alignment and  in smectic A clusters, clusters of molecules are generally oriented perpendicular to the layers  and these clusters  are large enough to restrict the motion of the colloidal particle in the perpendicular direction to the layer plane. Hence, we could conclude only in case of presmectic A clusters, the power of $\omega$ reduces to $1/2$.
Further, in $a_{m}$ and $a_{s}$ regimes,   $G''(\omega)$ vs $\omega$ in parallel and perpendicular directions fits with 
\begin{equation}
G''(\omega) = a * \omega + b*\omega^{n}* sin^{2}\left(\frac{c}{\omega} \right) 
\end{equation}
 and in $a(b)$ and $a_{c}$ regimes, $G''(\omega)$ follows  power law dependence with $\omega$.
In $a_{m}$ regime,
 \begin{eqnarray}
  G''_{\|} \simeq \omega^{0.94} \cr
  G''_{\|} \simeq \omega^{0.95}
 \end{eqnarray}

  In $a_{s}$ regime,
  \begin{eqnarray}
   G''_{\|} \simeq \omega^{0.74}\cr
  G''_{\|} \simeq \omega^{0.85}
  \end{eqnarray}

  Also, we computed the dynamic viscosity ($\eta''$) values from  $ G''/\omega$ in both parallel and perpendicular directions to the director $(\hat{\textbf{n}})$. The characteristics of ($\eta''$) are very similar to bulk rheology viscosity values\cite{K. Negita}. But, we observed a crossover between $\eta''_{\|}$ and $\eta''_{\bot}$. Similar crossover between $\eta_{\|}$ and $\eta\bot$ in the micro viscosity values of bent core liquid crystals near N-Sm-A transition were observed in our previous paper\cite{dhara}.

   \begin{figure}
\includegraphics[scale=0.32]{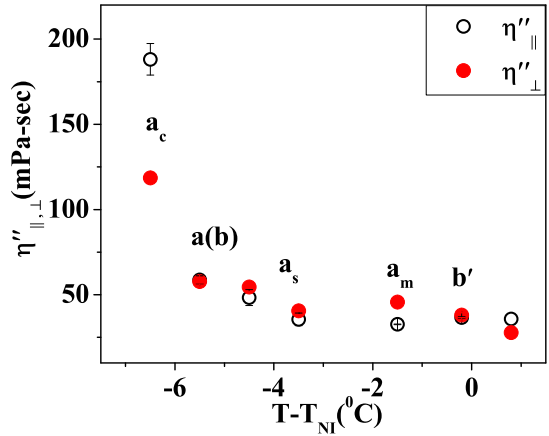}
\caption{Temperature dependent variation of dynamic viscosity($\eta^{''})$. $\eta_{\|}$ and $\eta_{\bot}$ are represented open  and closed circles(red colour online)respectively.}
\label{8CB viscosity}
\end{figure}
  
 Figure\ref{8CB viscosity} shows that as  $a_{c}$ approaches the smectic region, the thermal motion of the microsphere in the parallel directions(Y-direction) becomes highly restricted by more smectic layer formations. One can understand this restriction of thermal motion of microsphere as since the cell is rubbed to promote specfic molecular alignment, the larger smectic layer planes in $a_{c}$ regime are oriented perpendicular to the rubbing direction and are large enough to restrict the thermal motion of microsphere. Hence, the self diffusion of the microsphere in the direction perpendicular to the layer plane is lower than in parallel direction as a result $\eta''_{\|}$ can diverge.

\section{Conclusions}
	As per the  ELP theory of nemato dynamics, shear flow tilts the orientation of  $\hat{\textbf{n}}$ in liquid crystals. The results of bulk rheology measurements studied earlier in 8CB liquid crystals had always been explained in terms of these tilts \cite{C. R. Safinya}, \cite{D.F.Gu}, \cite{K. Negita}.  On lowering the temperature from N-Sm-A, the consequences of precessional motion of  $(\hat{\textbf{n}})$ along different axial directions had been observed. We have established three important phenomenon in 8CB liquid crystals, using particle tracking method.
	 
\begin{enumerate}
\item	Using passive microrheology measurements (zero shear), results of tumbling nematic director is observed in $G'$ and $G''$ measurements. 

\item Physical origin for the tumbling nematics in liquid crystals was therefore redefined for zero shear conditions.

\item $G'$ vs $\omega$ measurements were reported to  differentiate pure nematic phase in $b'$ regime,  smectic C like clusters  in the presmectics of $a_{m}$ and $a_{s}$ regimes and smectic A layers in the presmectics of $a(b)$ and $a_{c}$ regimes. 

We  have also shown in this paper dynamic viscosity measurements  which are in concurrence with previous shear rheology \cite{K. Negita}. This is the first evidence of the tumbling of the nematic director even under near-zero shear condition.  
\end{enumerate}
	 
\section{Acknowledgement}	
Y. Balaji thanks ACRHEM, University of Hyderabad for fellowship.

\end{document}